\documentclass{article}

    \PassOptionsToPackage{numbers, compress}{natbib}

\usepackage[final]{neurips_2021}

\usepackage[utf8]{inputenc} 
\usepackage[T1]{fontenc}    
\usepackage{hyperref}       
\usepackage{url}            
\usepackage{booktabs}       
\usepackage{amsfonts}       
\usepackage{nicefrac}       
\usepackage{microtype}      
\usepackage{xcolor}         
\usepackage{graphicx}
\usepackage{xcolor}
\usepackage{subcaption}

\definecolor{tagcolor}{RGB}{238,230,255}

\newcommand{\code}[1]{\texttt{#1}}

\title{Deep Learning Tools for Audacity: \\ Helping Researchers Expand the Artist's Toolkit}

\author{%
  Hugo Flores Garcia$^1$, Aldo Aguilar$^1$, Ethan Manilow$^1$, Dmitry Vedenko$^2$, Bryan Pardo$^1$ \\
  $^1$ Northwestern University, \quad $^2$ Audacity Team\\
  \texttt{hugofg@u.northwestern.edu} \\
}

\begin{document}

\maketitle


\begin{figure}[h]
    \centering
    \includegraphics[width=1\columnwidth]{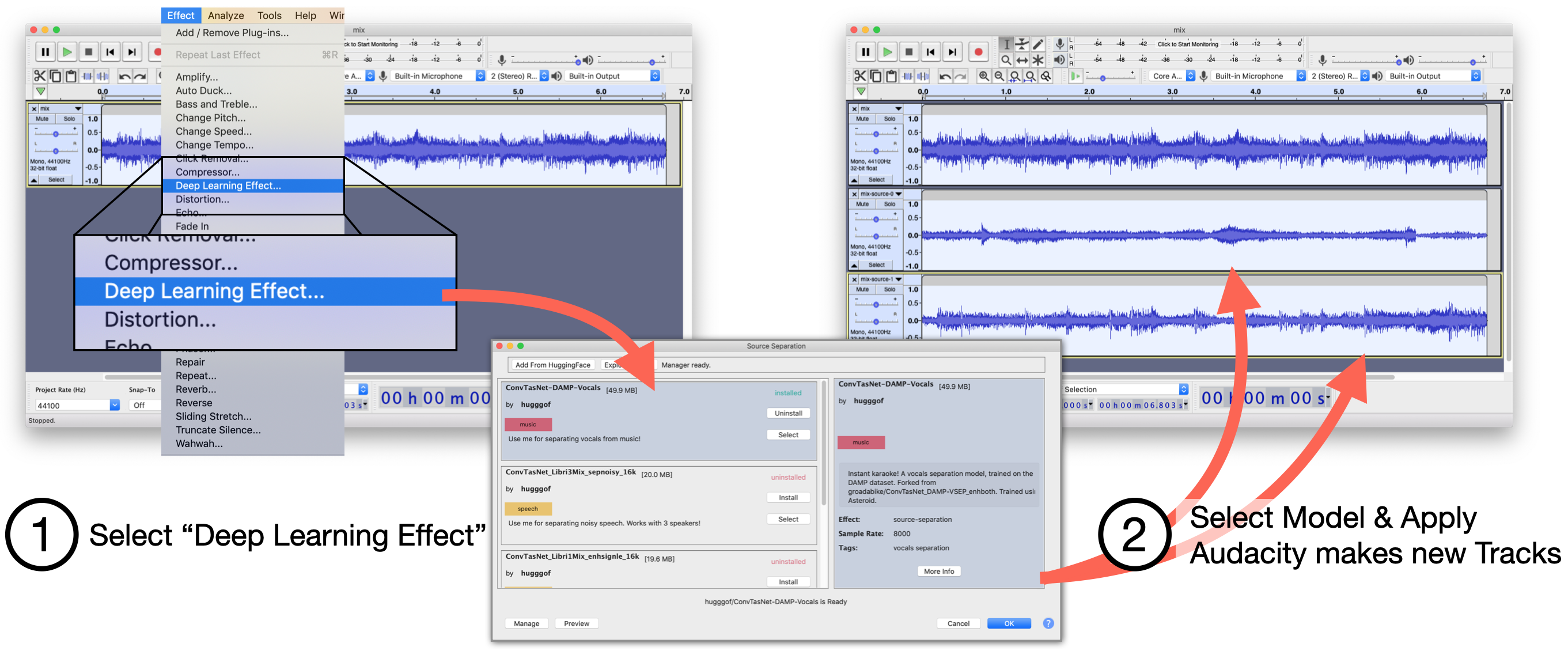}
    \caption{Extracting the vocals for later remixing using a deep source separation model in the Audacity Digital Audio Workstation (DAW). The sound artist selects the desired model available on HuggingFace. It automatically downloads and runs locally, removing the need to upload potentially private audio to a distant website or compile and run code to use a model locally.}
    \label{fig:example}
\end{figure}

\section{Introduction}
\label{sec:intro}

Digital Audio Workstations (DAWs) such as Audacity, ProTools and GarageBand are the primary platforms used in the recording, editing, mixing, and producing of sound art, including music. Making deep models for audio processing easily available on these platforms would be transformative for many artists. For example, remix artists (e.g. Public Enemy, Girl Talk, Kanye West) recombine elements of existing tracks into new works. This often requires spending many hours using a DAW to hand-separate individual vocals, drum beats, or other sounds out of recordings that contain multiple concurrent sounds. Deep learning researchers have created many effective models \cite{asteroid, manilow2020opensourceseparation, nussl} for separating out voice or individual instruments from music recordings, but these models are typically available only as source code on PapersWithCode or HuggingFace. Artists could save many hours of work, if only there were an easy way to add them into a DAW. More generally, deep models have greatly increased the set of audio generation and manipulation tools that are possible, including some that would be unimaginable or infeasible with traditional DSP techniques, such as melody co-creation \cite{vincentai}, automated upmixing \cite{jarnow2021upmix} and  replacing the timbre of one sound (e.g. human voice) in an existing audio recording with that of another (e.g. a violin) \cite{mor2018autoencoderbased, engel2020ddsp}. How can we best put these new tools in the hands of the artists?

 The standardization of modular effects (e.g. reverberation, equalization, compression) and synthesizer ``plugins'' has allowed independent developers of traditional DSP-based effects to easily add to the set of tools available to sound artists usind DAWs.  Unfortunately, the skills a person needs to develop a novel deep learning model are different than those required to build a plugin for a DAW. In practice, this means that deep learning technology rarely makes its way into DAW-hosted plugins. Furthermore, while it has become increasingly common for deep learning developers to provide runnable source code or even web demos of their work, most end-users do not have the skills to run open-sourced code. Even for those able to overcome this barrier, leaving the DAW to use research code or online demos can form a significant impediment to the workflow, discouraging use. Simplifying the deployment pipeline so that deep model developers can easily make models available to be used directly in the DAW could have a transformative impact on the range of creative tools available to sound artists, producers and musicians.


In this work, we describe a software framework that lets ML developers easily integrate new deep-models into Audacity, a free and open-source DAW that has logged over 100 million downloads since 2015 \cite{fosshubaudacity}. Developers upload their trained PyTorch \cite{paszke2019pytorch} model to HuggingFace's Model Hub. The model becomes accessible through Audacity's UI and loads in a manner similar to traditional plugins. This lets deep learning practitioners put tools in the hands of the artistic creators  without the need to do DAW-specific development work, without having to learn how to create a VST plugin, and without having to maintain a server to deploy their models.

\section{Model Lifecycle -- From Contributor to End User}

Our framework defines an API for two common deep learning-based audio paradigms, namely audio-in-audio-out tasks (such as source separation, voice conversion, style transfer, amplifier emulation, etc.) 
and audio-to-labels tasks (e.g. pitch-tracking, voice transcription, sound object labeling, music tagging).
We illustrate the utility of this by incorporating two example deep models into Audacity: a source separator and a musical instrument labeler (see the Appendix). These deep learning plugins are situated alongside traditional, DSP-based effects in the plugin drop down menu, providing a means for an end-user to run deep learning models locally without leaving Audacity (Figure \ref{fig:example}).

The model contribution process only requires developers to be familiar with PyTorch. Finished models are serialized with \code{torchscript}\footnote{\url{https://pytorch.org/docs/stable/jit.html}} such that they can be run on an end-user's machine. While \code{torchscript} provides a more limited functionality than full PyTorch, we find that it nonetheless provides most necessary utilities for serializing Audacity-bound models. Serialized models can then be uploaded to the HuggingFace\footnote{\url{https://huggingface.co/}} model hub. Models uploaded to HuggingFace with the special \colorbox{tagcolor}{\code{audacity}} tag will automatically be scraped, so that end-users can download and run the models through the Audacity interface (shown in Figure~\ref{fig:manager-scroller}). Further developer details are provided in the Appendix.

Once a model has been successfully uploaded to HuggingFace, an Audacity user can find it by selecting  ``Deep Learning Effect'' in the ``Effect'' menu (for audio-in-audio-out models), or ``Deep Learning Analyzer'' in the ``Analyze''  menu (for audio-to-label models). When the user selects one of these, a dialog box is shown that is populated with model metadata pulled from HuggingFace. This dialog box gives users the ability to browse models, download a desired model, and run it on selected tracks. The output of the model is either additional audio tracks (updates do not happen in-place) or audio labels, depending on the model selected.

\section{Conclusion}
This work has the potential to be transformational in increasing an artist's ability to access machine learning models for media creation. Researchers building the next generation of machine learning models for media creation will be able to put these models directly into the hands of the artists.
We hope that this work will foster a virtuous feedback loop between model builders and the artists that apply these models to create sound art.  The ability to quickly share models between model builders and artists should encourage a conversation that will deepen the work of both groups.

More details are provided in the Appendix and documentation for model contributors\footnote{\url{https://interactiveaudiolab.github.io/project/audacity}}.

\newpage

\section*{Ethical Implications of this Work}
\label{sec:ethics}
A small number of deep learning audio effects are available via web-demo or server-side upload. Those with privacy concerns about source material (e.g. artists working on unreleased works) may not be comfortable uploading their audio to a third party. Providing an easy way for artists to use models locally removes this concern, democratizing deep learning tools for audio manipulation in end-user devices. This work enables access to such tools for users who would otherwise have difficulty accessing them. As such, the tools described in this paper could foster an ecosystem of artists and deep learning practitioners. Importantly, this work also has the potential side effect of facilitating misuse by bad actors (e.g. putting tools that facilitate manipulating recorded speech in the hands of those that may wish to falsify evidence). Some may argue that access to deep audio manipulation tools should, therefore, be tightly controlled. We argue that the right solution is not to restrict artist access to any new editing tools that may arise. Instead, we encourage the exploration of solutions for detecting and addressing the unethical use of audio manipulation tools while preserving the open, unrestricted access to these systems for creative expression.


\section*{Acknowledgements}
\label{sec:ack}
This work was funded in part by a Google Summer of Code grant and by NSF Award Number 1901456. We would also like to thank Jouni Helminen and Prem Seetharaman for fruitful conversations.

\bibliographystyle{abbrv}
\bibliography{bibliography}



\appendix
\section*{Appendix}
\label{appendix}

\section{Examples}
\label{sec:examples}

We provide built-in models for several use cases: source separation (isolating vocals, or musical instruments from a recording),  voice enhancement, and musical instrument recognition.  In our supplementary video material~\footnote{\url{https://interactiveaudiolab.github.io/project/audacity}}, we showcase the creative opportunities allowed by our proposed framework by walking through the process of remixing a pop song, using different source separation models. 




\section{End user workflow}
\label{sec:enduser}

\begin{figure}[h]
    \centering
    \includegraphics[width=1\columnwidth]{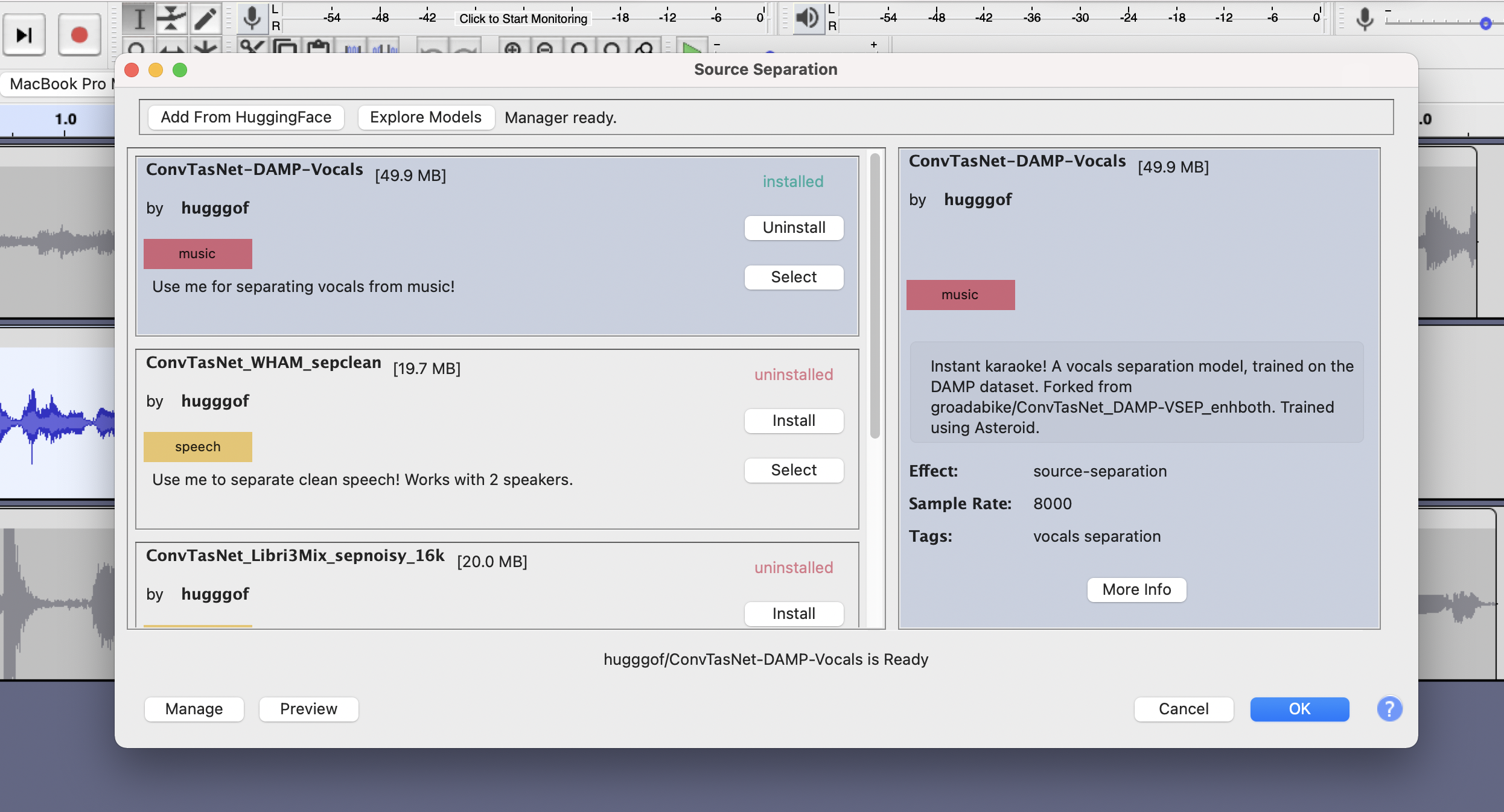}
    \caption{Model manager UI in Audacity. Users can explore a curated set of high-quality neural network models available on HuggingFace for a variety of use cases. Models can be contributed by anyone in the community. To see a full list of all models, the user clicks on ``Explore Models'', on the top left corner of the window.}
    \label{fig:manager-scroller}
\end{figure}

From an end-user perspective, Audacity offers two kinds of deep-learning tools, one for processing waveform-to-waveform (i.e. ``Deep Learning Effect'') and one for processing waveform-to-labels (i.e. ``Deep Learning Analyzer''). Users can process their audio by selecting a target audio region, followed by opening their desired deep learning tool. 
Deep Learning tools can be accessed via the ``Effect'' menu on Audacity's menu bar, while Analyzer tools can be accessed via the ``Analyze'' menu.
After installing and selecting a model, a user can apply the effect by pressing the ``OK'' button on the lower right, similar to other built-in offline effects in Audacity. Unlike other Audacity effects, however, the output audio is written to new tracks, instead of in-place. 

\begin{figure}[t]
    \centering
  \begin{subfigure}{.525\columnwidth}
    \includegraphics[width=1.0\columnwidth]{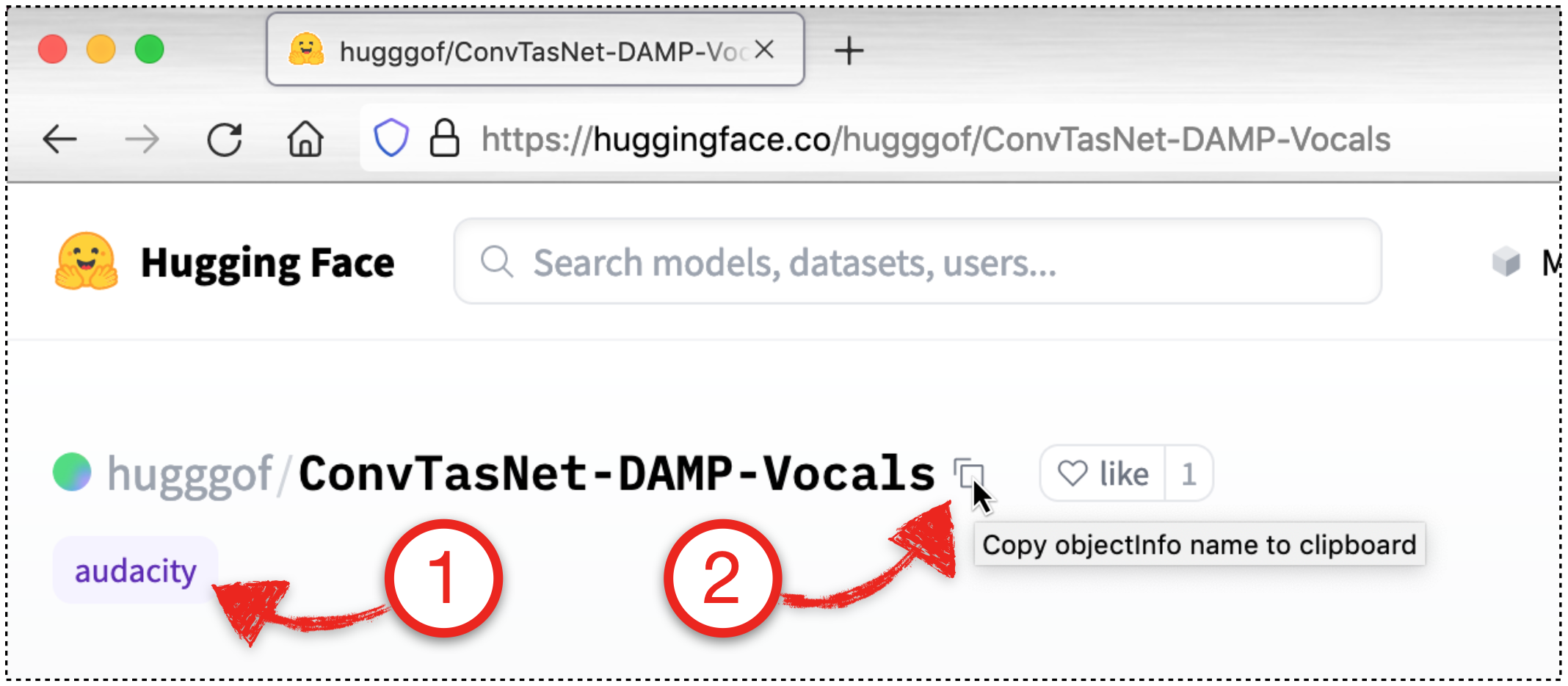}
    \caption{}
    \label{fig:huggingface-header}
    \end{subfigure}%
    \quad%
  \begin{subfigure}{.425\columnwidth}
    \centering
    \includegraphics[width=1.0\columnwidth]{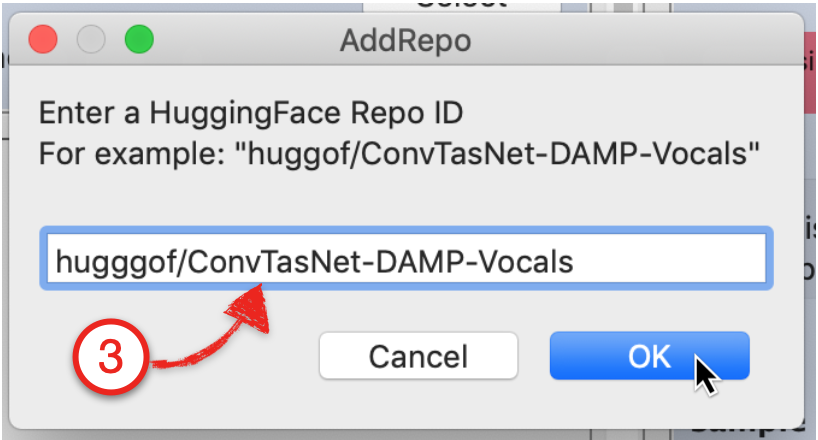}
    \caption{}
    \label{fig:addrepo}
    \end{subfigure}%
    \caption{Figure~\ref{fig:huggingface-header} (left) shows an annotated screenshot from the HuggingFace website showing a model that can be run with Audacity. For a model creator to make their model accessible to Audacity users, the model repository must have the \colorbox{tagcolor}{\code{audacity}} tag, denoted by the red ``1'' on the bottom left of Fig.~\ref{fig:huggingface-header}. From the HuggingFace website, users can click on the copy icon next to the repo's ID (denoted by the red ``2'') to copy the ID onto their clipboard. To add the model to Audacity, the user must paste the repo's ID on the dialog box shown on the right in Figure \ref{fig:addrepo} (shown by the red ``3'').}
\end{figure}

When a user opens either Deep Learning Effect or Analyzer for the first time, they are shown the model manager interface, shown in Figure \ref{fig:manager-scroller}. The model manager displays a curated list of repositories from HuggingFace, filtered by type (i.e. ``Effect'' or ``Analyzer''). The model manager interface resembles a package manager interface, providing a way to navigate through the list and install a model with a single button press. Additionally, users can click on the ``Explore Models'' button on the top toolbar to open a HuggingFace search query in an external browser for viewing/installing all models contributed to HuggingFace by the Audacity community. The user can add an uncurated model to their local collection by copying the repository's ID (e.g. \code{hugggof/ConvTasNet-Vocals}) into the ``Add from HuggingFace'' dialog, shown in Figure \ref{fig:addrepo}.  

Although the deep learning effects proposed here are applied off-line (i.e., not real-time), we find anecdotally that large models can perform inference quite quickly, even on commodity hardware. As an illustrative example, we used a mid-2014 MacBook Pro (7 years old at the time of this writing) to run the large source separation network Demucs~\cite{defossez2019music} ($\approx$ 265M parameters) on  210 sec (3:30 min) of monophonic audio recorded at a sample rate of 44.1 kHz (compact-disc quality audio). It took 80 seconds to process. This speed is in line with advanced DSP audio processing commonly applied by DAW users. Given that, we believe  this delay is acceptable for end-users, and this opens up the possibility of developing real-time deep learning effects in future work.

\section{Contributing models to Audacity}
\label{sec:contributing}

Deep learning models for Audacity are hosted in the HuggingFace model hub. The model contribution process is meant to be as straightforward as possible, requiring model contributors to be familiar with only PyTorch \cite{paszke2019pytorch}, a popular open-source and python-based deep learning framework. 

In order to make a PyTorch model compatible with our framework, a model contributor must 1) serialize their model into a \code{torchscript} file, 2) create a metadata file for the model, and 3) upload their model to the HuggingFace model hub. We provide an outline of the model contribution process in this work, as well as make thorough documentation for model contributors available online\footnote{\url{https://github.com/hugofloresgarcia/audacitorch}}. Additionally, we provide an example notebook\footnote{\url{https://github.com/hugofloresgarcia/audacitorch/blob/main/notebooks/example.ipynb}} that serializes a source separation model pre-trained using the Asteroid library~\cite{asteroid}. 

\subsection{Writing and Exporting Models for Audacity}
\label{sec:contributing-exporting}
After training a model using PyTorch, a contributor must serialize their model into the \code{torchscript} format to use in Audacity. We note that this poses limitations not present in pure Python environments, 
because the supported operations in \code{torchscript} are a subset of the wider Python and PyTorch operation set.
Nevertheless, we find that most necessary utilities for building end-to-end audio models are \code{torchscript} compatible: \code{torchscript} supports a large set of built-in functions and data types~\footnote{\url{https://pytorch.org/docs/stable/jit_builtin_functions.html}}, and the \code{torchaudio} package contains many serializable pre- and postprocessing audio modules (e.g., \code{Spectrogram}, \code{GriffinLim}, \code{MelSpectrogram}, Mel Filter Banks, etc.). 

\subsection{Model Architectures}
\label{sec:contributing-architectures}

\begin{figure}[h]
    \centering
    \includegraphics[width=1\columnwidth]{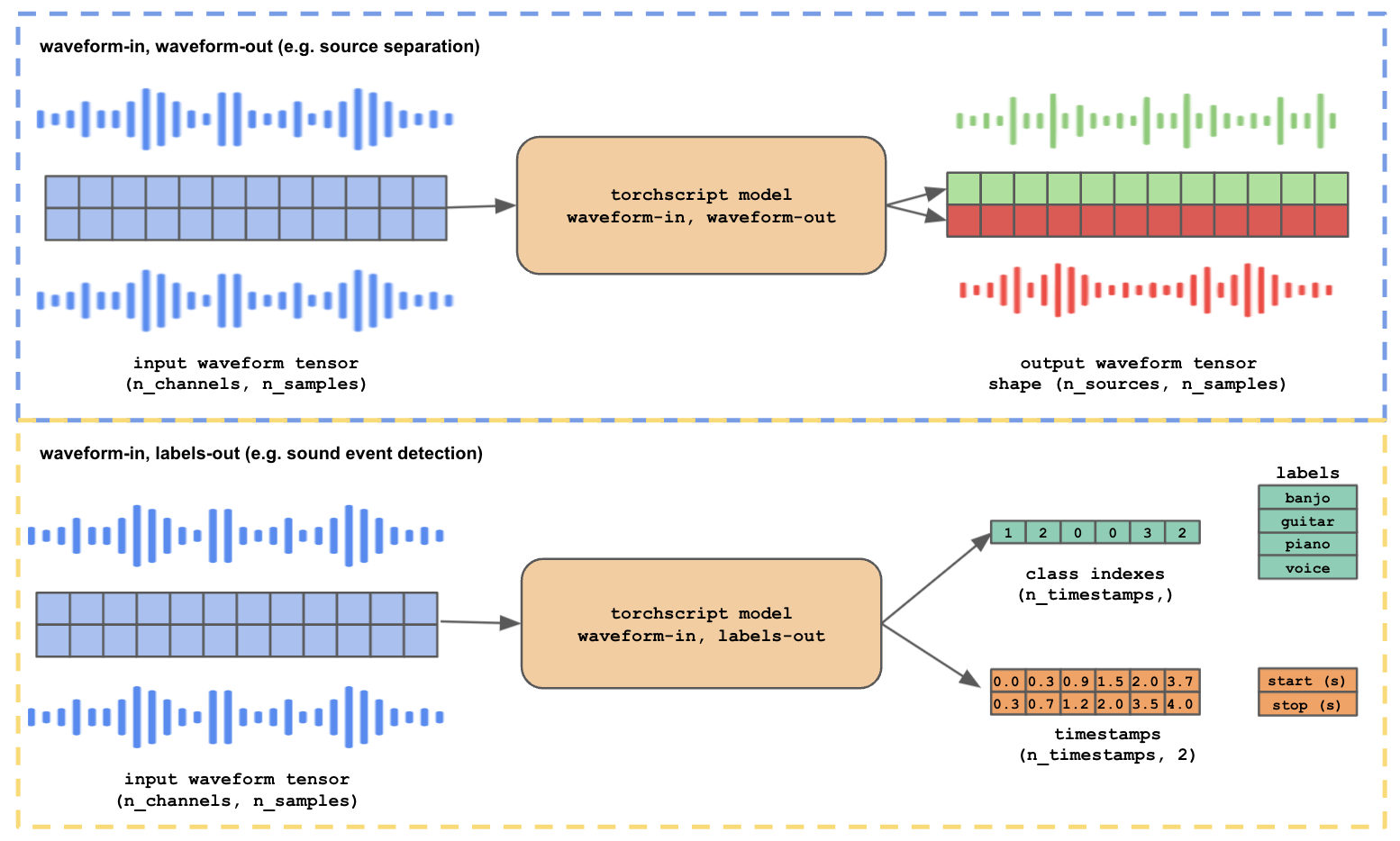}
    \caption{Tensor graphs for Audacity models: 1) waveform-to-waveform (top), and 2) waveform-to-labels (bottom). Note that the framework is agnostic to the internal model architecture, as long as it satisfies the input-output constraints, shown above. The \code{timestamps} tensor is shown transposed for compactness.}
    \label{fig:tensor-flow}
\end{figure}

As of this work, the deep learning framework for Audacity supports two kinds of deep learning-based effects: 1) waveform-to-waveform, and 2) waveform-to-labels. Both effects are \textbf{agnostic} to the internal model architecture, assuming the model is serializable via \code{torchscript} and satisfies Audacity's input-output constraints. These constraints are shown in Figure~\ref{fig:tensor-flow}, and outlined below:

\begin{itemize}
    \item During inference, both waveform-to-waveform and waveform-to-labels models are passed a raw audio multichannel waveform tensor with shape \code{(num\_input\_channels, num\_samples)} as input to the forward pass. 

    \item Waveform-to-waveform effects receive a single multichannel audio track as input, and may write to a variable number of new audio tracks as output. Example models for waveform-to-waveform effects include source separation, neural upsampling, guitar amplifier emulation, generative models, etc. Output tensors for waveform-to-waveform models must be multichannel waveform tensors with shape \code{(num\_output\_channels, num\_samples)}. For every audio waveform in the output tensor, a new audio track is created in the Audacity project. 
    
    \item Waveform-to-labels receive a single multichannel audio track as input, and may write to a single label track as output. The waveform-to-labels effect can be used for many audio analysis applications, such as voice activity detection, sound event detection, musical instrument recognition, automatic speech recognition, etc. The output for waveform-to-labels models must be a tuple of two tensors. The first tensor corresponds to the class indexes for each label present in the waveform, shape \code{(num\_timesteps,)}. The second tensor must contain timestamps with start and stop times for each label, shape \code{(num\_timesteps, 2)}.

\end{itemize}

\subsection{Model Metadata}
\label{sec:contributing-metadata}

Certain details about the model, such as its sample rate, tool type (e.g. waveform-to-waveform or waveform-to-labels), list of labels, etc. must be provided by the model contributor in a separate \code{metadata.json} file. In order to help users choose the correct model for their required task, model contributors are asked to provide a short and long description of the model, the target domain of the model (e.g. speech, music, environmental, etc.), as well as a list of tags or keywords as part of the metadata.

\subsection{Uploading to HuggingFace}
\label{sec:contributing-huggingface}

Once a model has been serialized to \code{torchscript} and a metadata file has been created, model contributors can upload their models to repositories in the  HuggingFace model hub. If the tag \colorbox{tagcolor}{\code{audacity}} is included in the repository's \code{README.md}, the repository will be visible to Audacity users via the ``Explore HuggingFace'' button (shown in Figure \ref{fig:manager-scroller}).

\end{document}